\documentclass[12pt]{article}
\usepackage{axodraw}
\usepackage{pstricks}
\usepackage{color}
\def\be{\begin{equation}}
\def\ee{\end{equation}}
\def\ba{\begin{eqnarray}}
\def\ea{\end{eqnarray}}

\begin{document}
\title{Cancellation of soft and collinear divergences in  noncommutative QED}
\author{B. Mirza\thanks{email: b.mirza@cc.iut.ac.ir},
 \ \ M. Zarei \thanks{email: zarei@ph.iut.ac.ir} \\ \\
{\it Department of  Physics, Isfahan University of Technology (IUT)}\\
{\it Isfahan,  Iran,} \\}

\date{}
\maketitle
\begin{abstract}
In this paper, we investigate the behavior of non-commutative IR
divergences and will also discuss their cancellation in the
physical cross sections. The commutative IR (soft) divergences
existing in the non-planar diagrams will be examined in order to
prove an all order cancellation of these divergences using the
Weinberg's method. In non-commutative QED, collinear divergences
due to triple photon splitting vertex, were encountered, which
are shown to be canceled out by the non-commutative version of
KLN theorem. This guarantees that there is no mixing between the
Collinear, soft and non-commutative IR divergences.
\end{abstract}

\newpage
\section{Introduction}
Non-commutative field theories have received a great deal of attention [1]. The non-commutativity  has also been
  studied phenomenologically in both high and low energy experiments [2].
These theories, however, suffer from certain shortcomings.
Field theories with space-time non-commutativity, for example, do not have a unitary $S$-matrix \cite{3}.
 The extra branch cuts, in theses theories, are developed in the loop diagrams which are responsible for the failure
 of the cutting
rules and lack of unitary $S$-matrix. It seems that we can save
unitarity at the expense of enlarging the Hilbert space of
asymptotic states, similar to making the $S$-matrix of open
string theory unitary by introducing the closed-string states
\cite{4}. The UV/IR mixing observed by Minwalla et al [5] appears
to be the main qualitative difference between conventional and
non-commutative perturbation field theories. In general, for any
non-commutative field theory, the loop diagrams can be classified
in the so-called planar and non-planar graphs. Perturbative
calculations have shown that the planar diagrams contain exactly
the same ultraviolet divergencies as their commutative
counterparts do. It has been found that there is an intriguing
mix of ultraviolet and infrared scales in the non-planar
diagrams. This is what is known as UV/IR mixing, which is a
general feature of the non-commutative field theories [1,5]. For
example, in the non-commutative $\Phi^{4}$ theory it has been
found that the effective action has a singularity at
$\widetilde{p}=0$ ($\widetilde{p}=p\theta$) that can be
interpreted as an IR divergence at fixed $\theta$ or as a
non-analytic behavior in the $\theta\rightarrow 0$ at fixed $p$
[5-7]. The UV/IR mixing in the case of gauge theories also shows
some specific features [8-12]. It has been shown that in the
non-commutative QED, a new IR divergence (non-commutative IR
divergence) appears \cite{6}. These non-commutative IR
divergences are different from IR (soft) divergences which are
related to the existence of massless photons in theory.
 It has long been  known that the soft divergences  occurring in QED can be canceled out
in transition rates or cross sections computed for detectors with finite energy resolution: the soft divergences
which occur in a scattering process due to the emission of an undetected soft real photons with total
energy $\leq E_{l}$ exactly cancel out the soft divergences due to virtual photon corrections order by order in
 perturbation theory. This cancellation was first shown by Bloch and Nordsieck \cite{13} in QED and is referred to as
Bloch-Nordsieck theorem. In ordinary (commutative) Yang-Mills theories just as in QED, there exist IR (soft) divergences
due to massless
gluons. In these theories, in addition to the soft divergences, the collinear divergences occur [14]. This kind of divergence is
due to the presence of triple gluon vertex. In external lines, gluon can decay into two collinear gluons that leads to collinear
divergences. In spite of this, it is believed
that the soft collinear divergencies in Yang-Mills theories are canceled out order by order in perturbation theory if the physical
cross section is calculated [15,16]. This has been guaranteed by the theorems of Kinoshita \cite{17} and of Lee and Nauenberg
\cite{18} known as KLN theorem which states that the transition rates are free of the collinear and soft divergencies
if we sum over initial and final states. This theorem
is a fundamental quantum mechanical theorem on the basis of unitarity of $S$-matrix. A great  many studies  have
investigated the cancellation
of collinear and soft divergences at one-loop and two-loop orders [19,20].
 In this paper the divergences  occurring in non-commutative
QED will be explored whose structure is the same as that in Yang-Mills theories. For our purposes, the one-loop
non-planar vertex correction will be investigated in order to identify
the soft and non-commutative IR divergences as well as the collinear divergence or mass singularity. It will be
shown that the soft divergences can be cancelled out by calculating the physical or measured cross section.
This method will be applied to the logarithmic non-commutative IR divergences appearing in vertex correction.
 Another aspect of the present study is to show that this kind of divergence is also canceled out in
the physical cross section if the energy resolution of detector is in the order of non-commutative parameter $\theta$.
 Next, we will turn to an all order soft divergence cancellation using Weinberg's method.
 This method will be applicable if the triple photon vertex is ignored. Otherwise, similar to Yang-Mills theories,
 the collinear divergences will appear up to the leading order in non-commutative QED. Therefore, it is
necessary to deal with the KLN theorem in the non-commutative space. This theorem is proved in the case where
unitarity is conserved ($\theta_{i0}=0$). Armed with this proof, the soft and collinear divergences will be shown to
 cancel out to all orders in the non-commutative QED.  Our
 motivation for considering KLN theorem in a non-commutative space
 was the following point: "In proceeding to higher loop orders,
 one faces the danger that the new IR divergences will mix with
 other divergences in an uncontrollable way, (in page 24 of Douglas and Nekerasov
 paper,[1])".

 This article is organized along the following lines. Section 2
presents the cancellation of soft and and non-commutative IR
divergences in vertex correction by computing the physical cross
section. Section 3 will provide the proof for the all order
 cancellation of soft divergences. Finally, in section 4  a non-commutative version of KLN theorem will be introduced
  and the cancellation of collinear divergences to all orders will be investigated.

\section{Soft and Non-commutative IR Cancellation in the Cross Section}
In this section, we will consider a fermion-fermion scattering in
non-commutative QED up to one loop order. The point here is  to
investigate the cancellation of soft and non-commutative IR
divergencies appearing in the non-planar diagrams in the physical
cross section.  Therefore, non-planar one-loop corrections will be
examined and the contribution of the cross section of the elastic
 scattering will be calculated with the emission of undetected soft photon
with an energy $\leq E_{l}$, the energy resolution of the
detector. Here an initial electron state $| e \rangle$ is
degenerate with the state $| e \gamma \rangle$ (an electron with
an emitted photon) when the photon is soft. The diagramatic
expression of the terms up to order
 $\alpha^{2}$ in the total cross section of the fermion-fermion scattering is shown in Fig.(1).

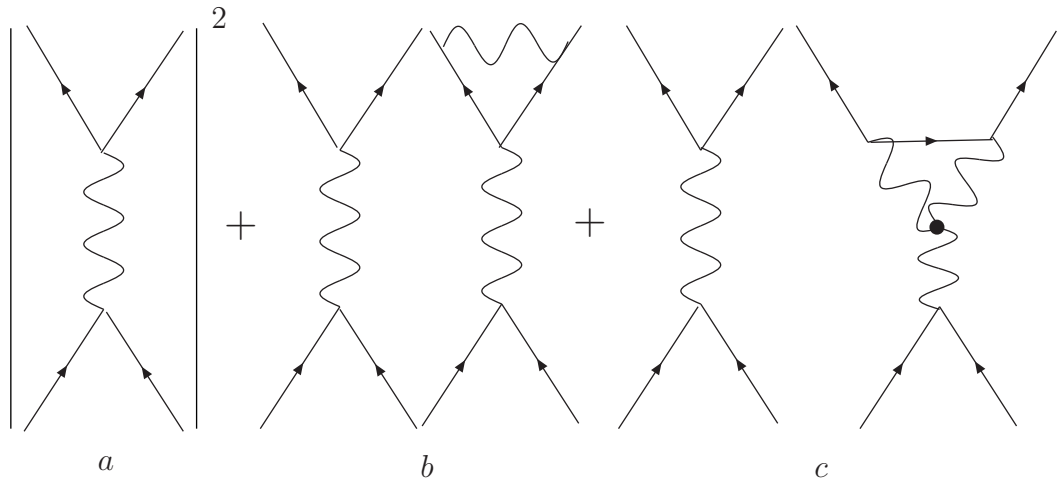
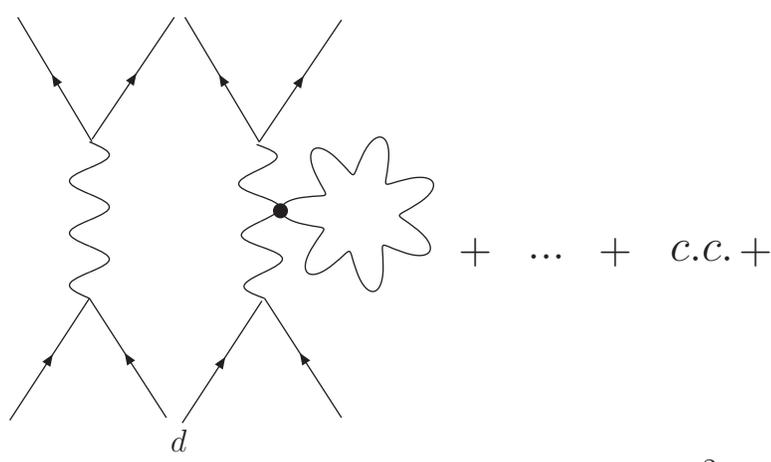
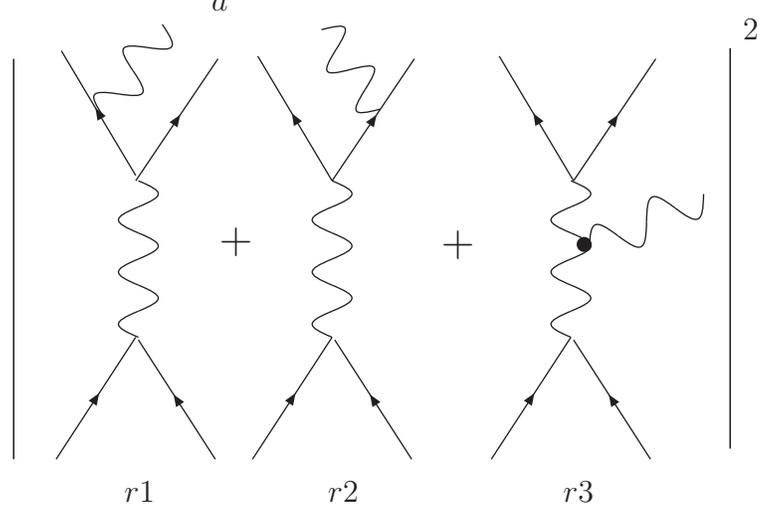
\begin{center}
\fcolorbox{white}{white}{
  \begin{picture}(405,545) (15,3)
    \SetWidth{0.5}
    \SetColor{Black}
    \GBox(26,379)(26,530){0.882}
    \GBox(96,379)(96,530){0.882}
    \Text(102,532)[lb]{\Black{$2$}}
    \Text(107,452)[lb]{\Large{\Black{$+$}}}
    \ArrowLine(31,378)(61,424)
    \Photon(61,483)(61,424){7.5}{3}
    \Photon(150,484)(150,425){7.5}{3}
    \ArrowLine(179,379)(150,424)
    \ArrowLine(150,484)(181,530)
    \Photon(211,485)(211,426){7.5}{3}
    \ArrowLine(211,485)(184,529)
    \ArrowLine(210,485)(241,531)
    \Photon(189,523)(236,525){7.5}{2}
    \Text(239,453)[lb]{\Large{\Black{$+$}}}
    \Photon(286,485)(286,426){7.5}{3}
    \ArrowLine(255,379)(285,425)
    \ArrowLine(286,484)(317,530)
    \Photon(376,424)(375,455){7.5}{2}
    \ArrowLine(349,487)(322,531)
    \ArrowLine(350,487)(396,488)
    \ArrowLine(395,488)(420,529)
    \ArrowLine(405,380)(376,425)
    \ArrowLine(346,379)(376,425)
    \Photon(376,455)(396,489){7.5}{2}
    \Vertex(375,455){2.83}
    \Photon(349,487)(377,455){7.5}{2}
    \ArrowLine(91,378)(62,423)
    \ArrowLine(120,379)(150,425)
    \ArrowLine(181,380)(211,426)
    \ArrowLine(240,381)(211,426)
    \ArrowLine(315,381)(286,426)
    \Photon(59,305)(59,246){7.5}{3}
    \ArrowLine(60,306)(91,352)
    \ArrowLine(60,483)(91,529)
    \ArrowLine(60,484)(32,531)
    \ArrowLine(149,485)(121,532)
    \ArrowLine(286,484)(258,531)
    \ArrowLine(60,305)(32,352)
    \ArrowLine(29,200)(59,246)
    \ArrowLine(88,201)(59,246)
    \ArrowLine(94,199)(124,245)
    \ArrowLine(123,305)(95,352)
    \ArrowLine(123,305)(154,351)
    \Photon(122,304)(125,246){7.5}{3}
    \Vertex(131,279){2.83}
    \ArrowLine(154,201)(125,246)
    \PhotonArc(161,278)(22.36,10,370){7.5}{7}
    \Text(199,259)[lb]{\Large{\Black{$+$}}}
    \Text(225,262)[lb]{\Large{\Black{$...$}}}
    \Text(305,259)[lb]{\Large{\Black{$+$}}}
    \Text(252,259)[lb]{\Large{\Black{$+$}}}
    \Text(279,261)[lb]{\Large{\Black{$c.c.$}}}
    \GBox(15,19)(15,170){0.882}
    \ArrowLine(31,19)(61,65)
    \ArrowLine(91,19)(62,64)
    \Photon(62,124)(62,65){7.5}{3}
    \ArrowLine(61,124)(92,170)
    \ArrowLine(61,126)(33,173)
    \Photon(49,148)(71,183){7.5}{2}
    \Text(93,96)[lb]{\Large{\Black{$+$}}}
    \ArrowLine(105,19)(135,65)
    \ArrowLine(165,19)(136,64)
    \Photon(135,124)(135,65){7.5}{3}
    \ArrowLine(135,124)(107,171)
    \ArrowLine(135,124)(166,170)
    \Photon(153,151)(131,181){7.5}{2}
    \Text(177,95)[lb]{\Large{\Black{$+$}}}
    \GBox(285,23)(285,174){0.882}
    \Text(291,178)[lb]{\Black{$2$}}
    \Photon(225,124)(225,65){7.5}{3}
    \ArrowLine(195,19)(225,65)
    \ArrowLine(255,19)(226,64)
    \Photon(232,98)(275,119){7.5}{2}
    \Vertex(230,100){2.83}
    \ArrowLine(226,124)(257,170)
    \ArrowLine(226,124)(198,171)
    \Text(181,362)[lb]{\Black{$b$}}
    \Text(57,3)[lb]{\Black{$r1$}}
    \Text(134,3)[lb]{\Black{$r2$}}
    \Text(223,3)[lb]{\Black{$r3$}}
    \Text(90,189)[lb]{\Black{$d$}}
    \Text(330,362)[lb]{\Black{$c$}}
    \Text(59,365)[lb]{\Black{$a$}}
  \end{picture}
}
\end{center}

Fig. 1: The diagramatic expression of the terms up to order
$\alpha^2$ in the total cross section of the fermion-fermion
scattering.

 Denoting the physical cross section to order
$\alpha^{2}$ due to non-planar diagrams, we have \be
\sigma_{ph}=\sigma_{a}+\sigma_{b}+\sigma_{c}+\sigma_{d}+\cdot\cdot\cdot+\sigma_{r}
\ee where $\sigma_{a},\sigma_{b},...$ are the cross sections
associated with diagrammatic expression represented in
Fig.(1a,1b,...) and the $\sigma_{r}$ denote the real-photon
emission cross section. It is known that
 non-planar diagrams are UV finite, hence no need for the field renormalization constant. Below is the  calculation
 of  amplitudes and the total cross section.

\subsection{Soft, Non-Commutative IR and Collinear Divergencies in the Vertex Correction}
The differential cross section in Fig. 1b  and its complex
conjugate is proportional to quantity $F$
 as defined by,
\be
F=\frac{e^{4}}{q^{4}}Tr[(p_{1}\!\!\!\!\!/+m)\gamma^{\mu}(p_{2}\!\!\!\!\!/+m)\gamma^{\nu}]Tr[(p_{1}\!\!\!\!\!/+m)(\Lambda_{\mu}+\Lambda^{\ast}_{\mu})(p_{2}\!\!\!\!\!/+m)\gamma_{\nu}]
\ee
Where $\Lambda_{\mu}+\Lambda^{\ast}_{\mu}$ is the one-loop non-planar vertex part and its complex conjugate. Using the Feynman rules
for non-commutative QED given in \cite{6} and \cite{21},  we will have,
\ba
&\Lambda_{\mu}+\Lambda^{\ast}_{\mu}& =\nonumber\\ &&=  -e^{2}\int\frac{d^{4}k}{(2\pi)^{4}}2\cos(k\times q)\frac{\gamma_{\sigma}}{k^{2}-\mu^{2}+i\epsilon}\frac{p'\!\!\!\!/-k\!\!\!/+m}{(p'-k)^{2}-m^{2}+i\epsilon}\gamma_{\mu}
\nonumber\\ && \times\frac{p\!\!\!/-k\!\!\!/+m}{(p-k)^{2}-m^{2}+i\epsilon}\gamma^{\sigma}
\ea
where $k\times q=k_{\mu}\theta^{\mu\nu} q_{\nu}$ and $m$ is the fermion mass and $\mu$ is
the fictitious mass for the photon to suppress the soft divergence.
Here, the non-commutative parameter $\theta^{\mu\nu}$ is assumed to be just magnetic i.e. $\theta^{i0}=0$. Applying
 Feynman parameterizatin, we will have,
\be
\Lambda_{\mu}+\Lambda^{\ast}_{\mu} = -e^{2}\int\frac{d^{4}k}{(2\pi)^{4}}2\cos(k\times q)\ \frac{numerator}{D^{3}}
\ee
where the denominator $D$ is,
\ba
& D & =(k^{2}-\mu^{2}+i\varepsilon)x  +((p'-k)^{2}-m^{2}+i\varepsilon)y+((p-k)^{2}-m^{2}+i\epsilon)z \nonumber\\ &&
     = k^{2}-2k\cdot (px+py)-z\mu^{2}-m^{2}(x+y)+xp'^{2}+yp^{2}+i\varepsilon
 \ea
Now shift $k$ to complete the square,
\be l\equiv k-xp'-yp \ee
After a bit of algebra, drawing on mass-shell condition and $q=p'-p$, we will obtain,
\be
D=l^{2}-\Delta+i\epsilon
\ee
where
\be
\Delta\equiv xyq^{2}-m^{2}(1-z)^{2}-\mu^{2}z
\ee
Based on the on-mass-shell condition and the anti-commutation relation, the numerator in (4) can be written as,
\ba
& Numerator & = \overline{u}(p)4[\gamma_{\mu}(2l^{2})-l\!\!/l_{\mu}]+p_{\mu}(4mx-4mxy-4my^{2})\nonumber\\ && +p'_{\mu}(4my-4mxy-4mx^{2})+ \gamma_{\mu}
            [2m^{2}(x+y)^{2}+m^{2}\nonumber\\ && -8m^{2}(x+y)+2q^{2}(x+y-xy-1)]u(p')
\ea
At this stage, we will focus on performing momentum integration and handling the terms that will contain the
 non-commutative IR divergence [6] i.e. the first term. Using the
formula given in \cite{22}, we can calculate the momentum integration. Below is the final result
including the divergent term,
\ba
& (\Lambda_{\mu}+\Lambda^{\ast}_{\mu})_{NC-IR} & =-e^{2}\int dxdydz\delta(x+y+z-1)\frac{4\pi^{2}}{2(2\pi)^{4}}\{(-2\gamma_{\mu})K_{0}(|\widetilde{q}|\sqrt{\Delta})\nonumber\\ &  &+ finite \ \ terms\}
\ea
where $|\widetilde{q}|=\sqrt{q_{\mu}\theta^{\mu\nu}\theta_{\nu\lambda}q^{\lambda}}$ and the subscript NC-IR mean the
non-commutative IR divergent part. In the $|\widetilde{q}|\rightarrow 0$ limit, we will  have a logarithmic divergence,
\ba
& (\Lambda_{\mu}+\Lambda^{\ast}_{\mu})_{NCIR} & =\frac{\alpha}{\pi}\gamma_{\mu}\int dxdydz\delta(x+y+z-1)\ln(|\widetilde{q}|\Delta)  +\nonumber\\ &  &  finite \ \ terms\}
\ea
Now let us consider the ordinary (commutative) IR divergence in (3). In the $\mu\rightarrow 0$ limit for the last
terms in (9) we have,
\ba
  & & -e^{2}\int dxdydz\delta(x+y+z-1) \frac{4\pi^{2}}{(2\pi)^{4}}\frac{1}{\Delta}\gamma_{\mu}\nonumber\\ &  &\times\left[2q^{2}(x+y-xy-1)
       +m^{2}(-8(x+y)+2(x+y)^{2}+1)\right]
\ea
Within this limit, the soft divergences occur in the corner of the Feynman parameter space where $z\approx 1$
(and, therefore, $x\approx y\approx 0$). In this region, we can set $z=1$ in the $\mu^{2}$ term in the denominator
 and for the sake of simplicity,  $z=1$ and $x=y=0$ in the numerators. Taking the $x$ integration by defining the new variables, the expression (12) will be simplified
to,
\be
  (\Lambda_{\mu}+\Lambda^{\ast}_{\mu})_{soft}=-\frac{\alpha}{2\pi}(-ie^{2})\log\left(\frac{-q^{2}}{m^{2}}\right)\log\left(\frac{-q^{2}}{\mu^{2}}\right)
\ee
There is no effect of non-commutative IR divergences in this region (the soft photon region) just as there are
no soft divergences in the noncommutative IR region. The cross section for process shown in Fig.(1a,b) for
non-planar diagrams and its complex conjugate is given as,
\be
 \left(\frac{d\sigma}{d\Omega}\right)_{total} =\left(\frac{d\sigma}{d\Omega}\right)_{NC-IR}+\left(\frac{d\sigma}{d\Omega}\right)_{soft}
\ee
which is equal to,
\be
  \left(\frac{d\sigma}{d\Omega}\right)_{B}\left[-\frac{\alpha}{\pi}\int dzdydz\delta(x+y+z-1)\ln(|\widetilde{q}|\sqrt{\Delta})-\frac{\alpha}{2\pi}
   \log\left(\frac{-q^{2}}{m^{2}}\right)\log\left(\frac{-q^{2}}{\mu^{2}}\right)\right]
\ee where $\left(\frac{d\sigma}{d\Omega}\right)_{B}$ denote the
Born cross section. Here, certain constant terms have been
ignored. In the next step, another divergence namely mass
singularity (collinear divergence) will be considered which
occurs when mass of the external particles is equal to zero. The
non-commutative IR divergence seems to be similar to mass
singularity. Taking the $m\rightarrow 0$ limit of (3) we get, \be
(\Lambda_{\mu}+\Lambda^{\ast}_{\mu})_{collinear}=-e^{2}\int\frac{d^{4}x}{(2\pi)^{4}}\frac{1}{k^{2}}
\frac{2\cos(k\times q)N_{\mu}}{(k^{2}-2k\cdot p')(k^{2}-2k\cdot
p)} \ee where $N_{\mu}$ is as defined in (9). With due attention
to the fact that, \be
\frac{1}{k^{2}}=\frac{1}{\omega}\left(\frac{1}{k_{0}-\omega+i\varepsilon}-\frac{1}{k_{0}+\omega-i\varepsilon}
\right) \ee with $\omega=|\textbf{k}|$ and performing the $k_{0}$
integration on complex $k_{0}$-plane, we have \ba
&(\Lambda_{\mu}+\Lambda^{\ast}_{\mu})_{collinear}=4\pi
e^{2}&\int_{0}^{\pi}d(\cos\theta)\int\frac{d\omega}{\omega}\cos(\omega\cdot
\widetilde{q}) \nonumber\\ &
&\times\frac{N'_{\mu}}{(p\,'_{\,0}-|\textbf{p}'|\cos\alpha)(p_{\,0}-|\textbf{p}|\cos\alpha)}
\ea where $\theta $ is the angle between $\textbf{p}'$ and
$\textbf{k}$ in the center of mass frame. In the $N'_{\mu}$,
$k_{0}$ has been replaced with $\omega$. In $m\rightarrow 0$
limit, in which we set the mass of external fermion to zero, we
have $p'_{\,0}=\textbf{p}'$ and $p_{\,0}=\textbf{p}$. So, \be
(\Lambda_{\mu}+\Lambda^{\ast}_{\mu})_{collinear}=4\pi
e^{2}\int\frac{d\omega}{\omega}\cos(\omega\cdot
\widetilde{q})\int_{0}^{\pi}d(\cos\theta)
\frac{N'_{\mu}}{1-\cos^{2}(\theta)} \ee Therefore, under the
condition where photon momentum $\textbf{k}$ is parallel to the
external momentum $\textbf{p}'$ ($\textbf{p}$), we face a
divergence in the angular integral or collinear divergence. This
kind of divergence is cancelled out in the physical cross section
[17,18].

\subsection{Divergence from Real-Photon-Emission}
 Now we turn to the calculation of the non-commutative photon emission differential cross section $d\sigma_{r}$.
 The non-planar diagramatics expression are shown in Fig.(1r). In any real experiment, a photon
detector can detect photon only down to some minimum energy resolution $E_{l}$. So in a real experiment, a scattered electron
can not be distinguished from an electron accompanied by a soft photon with energy down to energy resolution $E_{l}$. We
consider the diagrams for the emitted photon from external electron (figs.(1.r1,r2)). The calculation of $d\sigma_{r}$ follows,

\ba
& d\sigma_{r}& =d\sigma_{B}\int\frac{d^{3}k}{(2\pi)^{3}}\frac{1}{2k_{0}}\sum_{\lambda=1,2}e^{2}\left|\int-\frac{p'\cdot
\epsilon^{(\lambda)}}{p'\cdot k} e^{\frac{-i}{2}k\times
p'}+\frac{p\cdot \epsilon^{(\lambda)}}{p\cdot
k}e^{\frac{-i}{2}k\times p}\right|^{2}\nonumber\\ &&=
d\sigma_{B}\int\frac{d^{3}k}{(2\pi)^{3}}\frac{e^{2}}{2k_{0}} \nonumber\\ &&\times\left[-\frac{m^{2}}{(p'\cdot k)^{2}}-\frac{m^{2}}{(p\cdot k)^{2}}+\frac{2p\cdot
p'}{(p'\cdot k)(p\cdot k)}\cos\left(\frac{k\times
q}{2}\right)\right]
\ea

In the extreme relativistic limit, we consider only the third term. As already mentioned, the integration
is in the soft region: $0\leq k_{0}\leq E_{l}$. In order to make the integration well defined, we introduce a
 very small fictitious  mass $\mu$ for the photon which plays the role of a lower cutoff for the integral. So the expression (20) is simplified to,
\be
d\sigma_{r}=d\sigma_{B}\frac{\alpha}{\pi}\int_{\mu}^{E_{l}}\frac{dk_{0}}{k_{0}}I\cos(\frac{k\times q}{2})
\ee
where
\be
I=\int\frac{d\Omega}{4\pi}\frac{|\textbf{k}|^{2}(2m^{2}-q^{2})}{(k\cdot
p')(k\cdot p)}
\ee
Note that the quantity $I$ is independent of $k_{0}$; so, by performing the solid angle integration,
 in the high energy limit
$q^{2}\rightarrow \infty$, we obtain,
\be
I=2\log(\frac{-q^{2}}{m^{2}})
\ee
Therefore, the cross section $ \left(\frac{d\sigma}{d\Omega} \right)_{R}(p\rightarrow p'+\gamma(k<E_{l})) $, will  read as,
\be
 \left(\frac{d\sigma}{d\Omega} \right)_{r}=\left(\frac{d\sigma}{d\Omega} \right)_{B}\frac{\alpha}{2\pi}\left(\log\frac{E^{2}_{l}}{\mu^{2}}\log\frac{-q^{2}}{m^{2}}\right)
\ee
The non-commutative parameter  disappears when the integration is performed in the soft region. The cancellation
 of soft divergence in the vertex correction Fig(1.b) is also to be examined. As discussed earlier, on the scale
  of ordinary (commutative) QED, the cross section in a scattering event measured by the detector with a specific energy
resolution, is given by,
\ba
&\left(\frac{d\sigma}{d\Omega}\right)_{measured}&=\left(\frac{d\sigma}{d\Omega}\right)(p\rightarrow p')+\left(\frac{d\sigma}{d\Omega}\right)(p\rightarrow p'+\gamma(k<E_{l}))
\nonumber\\ &&=\left(\frac{d\sigma}{d\Omega}\right)_{r}+\left(\frac{d\sigma}{d\Omega}\right)_{soft}\nonumber\\ &&=
\left(\frac{d\sigma}{d\Omega}\right)_{B}\left[-\frac{\alpha}{2\pi}\log\left(\frac{-q^{2}}{m^{2}}\right)\log\left(\frac{-q^{2}}{\mu^{2}}\right)
+\frac{\alpha}{2\pi}\log\left(\frac{-q^{2}}{m^{2}}\right)\log\left(\frac{-E_{l}^{2}}{\mu^{2}}\right)\right]\nonumber\\ &&=
\left(\frac{d\sigma}{d\Omega}\right)_{B}\left[ \frac{\alpha}{2\pi}\log\left(\frac{-q^{2}}{E_{l}^{2}}\right)\log\left(\frac{-q^{2}}{m^{2}}\right)\right]
\ea
We see that the soft divergence has been canceled out in this process, which is in agreement with the result from [23].
This is a reliable method that works very well on this scale.

 In what follows, this method will be applied to the non-commutative region. In a non-commutative theory, for example
  non-commutative QED, we investigate the theory on non-commutative scales. So it will be reasonable to set the energy
   resolution of detector $E_{l}$ in the order of non-commutative parameter $\theta$.
We define the "non-commutative soft photon region" as
 $0\leq k_{\,0}=\sqrt{|\textbf{k}|^{2}+\lambda^{2}} \leq E_{l}\sim \omega^{2}|\widetilde{q}|$,
where $\omega$ is a constant with dimension $(energy)^{2}$. In this non-commutative region, a single-electron state
can not be distinguished from an electron state accompanied by the soft photon.
So for the cross section of the photon emitted from the external electron we have,
\be
d\sigma_{NC-r}=d\sigma_{B}\frac{\alpha}{\pi}\int_{\mu}^{\omega^{2}|\widetilde{q}|}\frac{dk_{0}}{k_{0}}I\cos\left(\frac{k\times k}{2}\right)
\ee
The remaining integration is performed in the $\widetilde{q}\rightarrow 0$ and nonrelativistic limit
and the final result of the cross section is obtained as follows,
\be
\left(\frac{d\sigma}{d\Omega}\right)_{NC-r}=\left(\frac{d\sigma}{d\Omega}\right)_{B}\frac{\alpha}{\pi}\,\ln(\frac{|\widetilde{q}|\omega^{2}}{\mu})
\ee
and therefore,
\be
\left(\frac{d\sigma}{d\Omega}\right)_{measured}\sim \, finite.
\ee
in which the measured cross section is free of both soft and non-commutative IR divergences. So
because of the existence of non-commutative soft photon, no non-commutative IR divergences
of the vertex correction are observed. The quadratic and logarithmic non-commutative
IR divergences that comes from new IR non-Abelian vertex and vacuum polarization corrections (Figs.(1c,1d,...)),
 may be cancelled out using other methods. For example, one may use the infrared rearrangement method [24] to
 cancel these non-commutative divergences. This
method is used in scalar field theories, where the IR divergences will appear
 when masses and external momenta are all equal to zero.
The non-commutative IR divergences are similar to this kind of divergence.

\section{The Cancellation of the Infrared Divergence to All Orders}
The structure of non-commutative QED is the same as Yang-Mills theories. We expect that at least in non-commutative space
($\theta_{i0}=0$) where the unitarity is confirmed, the soft and collinear divergences are cancelled out to all orders as
do in the Yang-Mills theories.
In ordinary QED, one approach is the cancellation of soft divergence in on-shell renormalization scheme in which the
 soft divergence
in $Z_{1}$, the vertex rescaling factor cancels against another from $Z_{2}$, the field-strength renormalization constant
as a result of the renormalization constant Ward identity ($Z_{1}=Z_{2}$) \cite{26}.
The result is general and can be extended to all orders. This  method is not appropriate in non-commutative QED
since in this theory $Z_{1}\neq Z_{2}$ \cite{27}.
However,  we will present an all-order proof of the cancellation using Weinberg's
lucid method \cite{28}. In this part, the non-planar
 vertex correction and the terms with the largest logarithmic enhancement at each
 order of perturbation theory will be discussed. It should be noted that the
 soft divergences from the other non-Abelian diagrams are
 finite.
Let us first consider the outcoming electron line with $n$ soft
photon attaching to it as shown in Fig. 2

\begin{center}
\fcolorbox{white}{white}{
  \begin{picture}(146,240) (210,-123)
    \SetWidth{0.5}
    \SetColor{Black}
    \GOval(290,-40)(12,12)(0){0.882}
    \Photon(264,8)(319,37){7.5}{3}
    \Photon(284,-51)(270,-103){7.5}{3}
    \Photon(296,-51)(328,-93){7.5}{3}
    \Vertex(287,5){1}
    \Vertex(295,-15){1}
    \Vertex(291,-5){1}
    \Text(326,38)[lb]{\Black{$k_{3}$}}
    \Text(310,63)[lb]{\Black{$k_{2}$}}
    \Text(296,91)[lb]{\Black{$k_{1}$}}
    \Photon(252,34)(304,62){7.5}{3}
    \ArrowLine(297,-31)(323,9)
    \ArrowLine(283,-30)(221,96)
    \Text(210,101)[lb]{\Black{$p_{2}$}}
    \Photon(240,57)(289,87){7.5}{3}

  \end{picture}
}
\end{center}

Fig. 2: The n soft photons attached to an outcoming electron.

\

The amplitude of this diagram is, \ba &\sum_{perm}&
\overline{u}(p_{2})(-ie\epsilon_{1})e^{\frac{i}{2}k_{1}\times
p_{2}}\frac{i(p_{2}\!\!\!\!\!/+k_{2}\!\!\!\!\!/+m)}{2p_{2}\cdot
k_{1}} \nonumber\\
&&\times(-ie\epsilon_{2})e^{\frac{i}{2}(k_{2}\times
p_{2}+k_{2}\times
k_{1})}\frac{i(p_{2}\!\!\!\!\!/+k_{1}\!\!\!\!\!/+k_{2}\!\!\!\!\!/+m)}{2p_{2}\cdot(
k_{1}+k_{2})}\cdot\cdot\cdot \nonumber\\
&&\times(-ie\epsilon_{n})e^{\frac{i}{2}(k_{n}\times
p_{2}+k_{n}\times(k_{1}+...+k_{n-1}))}\frac{i(p_{2}\!\!\!\!\!/+k_{1}\!\!\!\!\!/+...+k_{n}\!\!\!\!\!/+m)}{2p\cdot(k_{1}+\cdot\cdot\cdot+k_{n})}(iT_{0})
\nonumber\\ &&\simeq
(e)^{n}\sum_{perm}e^{\frac{i}{2}[(k_{1}+...+k_{n})\times
p_{2}+k_{2}\times
k_{1}+k_{3}\times(k_{1}+k_{2})+...+k_{n}\times(k_{1}+...+k_{n-1})]}
 \nonumber\\ &&\times\frac{p_{2}\cdot \epsilon_{1}}{p_{2}\cdot k_{1}}\frac{p_{2}\cdot \epsilon_{2}}{p_{2}\cdot (k_{1}+k_{2})}\cdot\cdot\cdot \frac{p_{2}\cdot \epsilon_{n}}{p_{2}\cdot ( k_{1}+...+k_{n})}\overline{u}(p_{2})(iT_{0})
\ea
where $\epsilon_{i}=\epsilon(k_{i})$ is the polarization with $i=1,2,...,n$ and by "perm" we mean all the
possible permutations
of index $(1,2,...,n)$ . Applying the Eikonal approximation, we assume that all the $k_{i}s$ are soft and so drop
the $O(k^{2})$
terms in the denominators and the $k_{i}$ in the numerator. We must hold the exponential in an all-order treatment;
 however, the terms like $k_{i}\times k_{j}$
can be ignored (compared to terms $p_{2}\times k_{i}$ in the
exponential). The sum of permutation can be performed by means of
the following formula, \be \sum_{perm}\frac{1}{p\cdot
k_{1}}\frac{1}{p\cdot(k_{1}+k_{2})}\cdot\cdot\cdot\frac{1}{p\cdot(k_{1}+...+k_{n})}=
\frac{1}{p\cdot k_{1}}\frac{1}{p\cdot
k_{2}}\cdot\cdot\cdot\frac{1}{p\cdot k_{n}} \ee The proof of this
formula follows from mathematical induction on $n$. Applying (30)
to (29), we find the coefficient of $(iT_{0})$ as \be
\overline{u}(p_{2})e^{n}\frac{p_{2}\cdot \epsilon_{1}}{p_{2}\cdot
k_{1}}\frac{p_{2}\cdot \epsilon_{2}}{p_{2}\cdot
k_{2}}\cdot\cdot\cdot\frac{p_{2}\cdot \epsilon_{n}}{p_{2}\cdot
k_{n}}e^{\frac{i}{2}p_{2}\times(\sum_{i=1}^{n}k_{i})} \ee A
similar set of manipulations simplifies the sum over soft photon
insertion on the incoming electron line (Fig. 3),
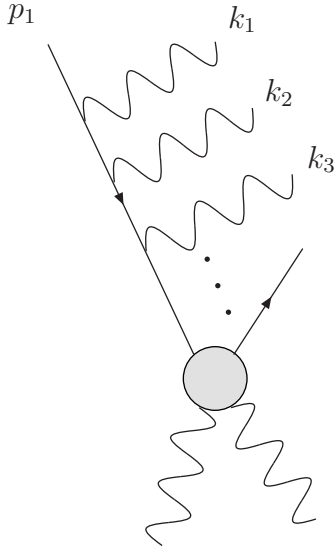
\begin{center}
\fcolorbox{white}{white}{
  \begin{picture}(143,235) (213,-130)
    \SetWidth{0.5}
    \SetColor{Black}
    \GOval(290,-45)(12,12)(0){0.882}
    \Photon(264,3)(319,32){7.5}{3}
    \Photon(284,-56)(270,-108){7.5}{3}
    \Photon(296,-56)(328,-98){7.5}{3}
    \Vertex(287,0){1}
    \Vertex(295,-20){1}
    \Vertex(291,-10){1}
    \Text(326,33)[lb]{\Black{$k_{3}$}}
    \Text(310,58)[lb]{\Black{$k_{2}$}}
    \Text(296,86)[lb]{\Black{$k_{1}$}}
    \Photon(241,52)(290,82){7.5}{3}
    \Photon(252,29)(304,57){7.5}{3}
    \ArrowLine(297,-36)(323,4)
    \ArrowLine(227,81)(282,-36)
    \Text(213,89)[lb]{\Black{$p_{1}$}}
  \end{picture}
}
\end{center}

Fig. 3: The n soft photons attached to an incoming electron.

\

 In this case, we get an extra minus sign in the factor for each
photon, since in the denominator we
 have $(p_{1}-\sum k)^{2}-m^{2}\approx -2p\cdot(\sum k)$. Also in the exponential, we have an extra
minus sign. It is now straightforward to obtain the replacement
rule for $n$ soft photons attached to $N$ electron lines with all
possible partitions $(n_{1},n_{2},...,n_{N})$ of soft photons
with the constraint $\sum_{i=1}^{N}=n$, as shown by a typical
diagram in Fig. 4,

\begin{center}
\fcolorbox{white}{white}{
  \begin{picture}(168,229) (119,-122)
    \SetWidth{0.5}
    \SetColor{Black}
    \GOval(207,2)(15,15)(0){0.882}
    \ArrowLine(216,-11)(241,-69)
    \ArrowLine(209,-13)(215,-86)
    \ArrowLine(203,-13)(196,-50)
    \ArrowLine(196,-9)(164,-73)
    \ArrowLine(197,14)(168,79)
    \ArrowLine(205,17)(186,87)
    \ArrowLine(212,17)(220,53)
    \ArrowLine(218,13)(249,82)
    \Photon(188,79)(224,102){7.5}{2}
    \Photon(193,63)(236,84){7.5}{2}
    \Photon(236,50)(287,66){7.5}{3}
    \Photon(181,-37)(129,-51){7.5}{3}
    \Photon(169,-64)(119,-75){7.5}{3}
    \Photon(214,-72)(164,-94){7.5}{3}
    \Photon(234,-53)(275,-33){7.5}{2}
    \Photon(180,52)(131,33){7.5}{3}
    \Photon(243,68)(278,88){7.5}{2}

  \end{picture}
}
\end{center}

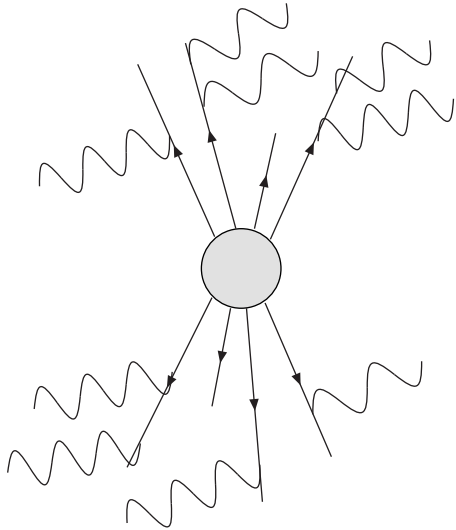
Fig. 4: A typical diagram with n soft photons attached to N
external electrons.

\

 The extra factor for this attachment is \be
\sum_{i=1}^{N}\prod_{l=1}^{n_{i}}e\eta_{i}\frac{p_{i}\cdot\epsilon_{l}}{p_{i}\cdot
k_{l}}e^{\frac{-i}{2}p_{i}\times\sum_{j=1}^{n_{i}}k_{j}} \ee

where $\eta$ is the signature factor defined by $\eta= \pm 1 $ for
incoming and outcoming electrons respectively.

  The non-commutative transition
amplitude with $N$ external electrons and $n$ soft photons can be
written in the following form \be
T_{n}=T_{0}\prod_{l=1}^{n}e\eta_{i}\frac{p_{i}\cdot\epsilon_{l}}{p_{i}\cdot
k_{l}}e^{\frac{-i}{2}p_{i}\times\sum_{j=1}^{n_{i}}k_{j}} \ee
$T_{0}$ is the transition amplitude without photon line in the
non-commutative space. The transition rate $\omega_{n}$ for
emitting $n$ photons in the non-commutative QED is given by \be
\omega_{n}=\frac{1}{n!}\int_{R}(\prod_{l=1}^{n}\frac{d^{3}k_{l}}{(2\pi)^{3}2k_{l0}})\sum_{pol}|T_{n}|^{2}
\ee where $\sum_{pol}$ means sum over polarizations and is
performed using  $-g_{\mu\nu}$ in place of
$\sum\varepsilon_{\mu}\varepsilon_{\nu}^{\ast}$. The soft photon
region $R$ is defined as

\be R=\{k_{1},...,k_{n};\sum_{l=1}^{n}k_{l0}\leq  E_{l} \} \ee

where $E_{l}$ is the energy resolution of the detector. After
summing the polarization, we obtain, \be
\omega_{n}=\frac{1}{n!}|T_{0}|^{2}\int_{R}\prod_{l=1}^{n}\{\frac{d^{3}k_{l}}{(2\pi)^{3}2k_{l0}}\sum_{i,j}e^{2}\eta_{i}\eta_{j}\frac{-p_{i}\cdot
p_{j}}{(p_{i}\cdot k_{l})(p_{j}\cdot
k_{l})}\}e^{\frac{i}{2}\widetilde{Q}_{ij}\cdot\sum_{r=1}^{n}k_{r}}
\ee with
$\widetilde{Q}_{ij}=\widetilde{p_{i}}-\widetilde{p}_{j}$. For
example, when the number of external electrons is $N=2$, we have
$\widetilde{Q}_{ij}=\widetilde{q}$, the momentum transfer. The
physical transition rate $\omega$ for $N$-electron process is
given by adding all orders, \ba
&\omega&=\sum_{n=0}^{\infty}\omega_{n}\nonumber\\
&&=|T_{0}|^{2}\sum_{n=0}^{\infty}\frac{1}{n!}\sum_{i,j=1}^{N}f
\nonumber\\ &
&\times\int\frac{dk_{10}}{k_{10}}e^{\frac{i}{2}|\widetilde{\textbf{Q}}_{ij}|k_{10}\cos(\alpha_{1})}\cdot\cdot\cdot
\frac{dk_{n0}}{k_{n0}}e^{\frac{i}{2}|\widetilde{\textbf{Q}}_{ij}|k_{n0}\cos(\alpha_{n})}\Theta(E_{l}-\sum_{r=1}^{n}k_{r0})
\ea where $\alpha_{i}$ is the angle between $ \textbf{k}_{ij}$
and $\widetilde{\textbf{Q}}_{ij}$. The step function $\Theta$ is
defined as, \be \Theta(\omega-\omega')=\frac{-1}{2\pi
i}\int_{-\infty}^{\infty}dx\frac{e^{-i(\omega-\omega')x}}{x+i\epsilon}=\frac{1}{\pi}\int_{-\infty}^{\infty}dx\frac{sin\omega
x}{x+i\epsilon}e^{i\omega'x} \ee and \be
f=\int\frac{d\Omega}{2(2\pi)^{3}}e^{2}\eta_{i}\eta_{j}\frac{-p_{i}\cdot
p_{j}}{(p_{i}\cdot k)(p_{j}\cdot k)}k_{0}^{2} \ee Here $d\Omega$
is the $\theta,\phi $ are angles describing $\overrightarrow{q}$.
By applying Feynman parameterization, it is easy to confirm that
the quantity $f$ is independent of $k_{0}$ and is a function of
$p_{i}\cdot p_{j}$. It is straightforward to see that the
quantity $ \widetilde{\textbf{Q}}_{ij}\cos(\alpha)$ will be
cancelled out when we take the integration over $k_{l0}$,

\ba
& \int_{\mu}^{E_{l}}\frac{dk_{0}}{k_{0}}e^{ik_{0}x}e^{\frac{i}{2}|\widetilde{\textbf{Q}}_{ij}|k_{0}\cos(\alpha)}&=\int_{\mu}^{E_{l}}\frac{dk_{0}}{k_{0}}e^{ik_{0}(x+\frac{1}{2}+|\widetilde{\textbf{Q}}_{ij}|\cos(\alpha))}\nonumber\\
&& \approx
\ln(k_{0}(x+\frac{1}{2}|\widetilde{\textbf{Q}}_{ij}|\cos(\alpha)))|_{\mu}^{E_{l}}\nonumber\\
&&=\ln \frac{E_{l}}{\mu} \ea The physical transition rate for
emission of any number of soft photons is therefore, \be
\omega=|T_{0}|^{2}\frac{1}{\pi}\int_{-\infty}^{\infty}dx\frac{\sin
x}{x+i\varepsilon}\exp\left(f\ln(\frac{E_{l}}{\mu})\right)=|T_{0}|^{2}\left(\frac{E_{l}}{\mu}\right)^{f}
\ee Let us now consider the situation in which there is no
external photon line. We turn to the virtual photon line
 participating only as an internal line in $T_{0}$. Denoting by $t_{n}$ the transition amplitude of $N$-electron
 with insertion of $n$ internal photon lines (Fig. 5), we have,
\be
T_{0}=\sum_{n=0}^{\infty}t_{n}\\ + \\\ finite \\\ terms
\ee
\begin{center}
\fcolorbox{white}{white}{
  \begin{picture}(258,168) (116,-172)
    \SetWidth{0.5}
    \SetColor{Black}
    \ArrowLine(145,-65)(127,-12)
    \ArrowLine(150,-65)(150,-11)
    \ArrowLine(155,-66)(173,-14)
    \ArrowLine(153,-91)(167,-146)
    \ArrowLine(146,-91)(133,-147)
    \ArrowLine(142,-89)(119,-143)
    \ArrowLine(157,-89)(178,-142)
    \ArrowLine(150,-91)(150,-149)
    \ArrowLine(140,-69)(116,-18)
    \Text(177,-83)[lb]{\Large{\Black{$=$}}}
    \Text(193,-84)[lb]{\Large{\Black{$\sum$}}}
    \Vertex(227,-80){12.04}
    \ArrowLine(218,-72)(197,-16)
    \ArrowLine(226,-72)(226,-4)
    \Photon(204,-36)(250,-36){7.5}{2}
    \ArrowLine(235,-74)(259,-17)
    \ArrowLine(220,-85)(196,-143)
    \ArrowLine(227,-92)(227,-153)
    \ArrowLine(235,-89)(259,-144)
    \Photon(198,-138)(257,-137){7.5}{3}
    \ArrowLine(159,-68)(184,-19)
    \PhotonArc(209.83,-76.67)(64.25,-50.16,46.58){-7.5}{5.5}
    \Text(297,-80)[lb]{\Black{$+$}}
    \GOval(150,-78)(13,12)(0){0.882}
    \Text(144,-82)[lb]{\Black{$T_{0}$}}
    \Text(316,-80)[lb]{\Black{$finite$}}
    \Text(352,-78)[lb]{\Black{$terms$}}
  \end{picture}
}
\end{center}
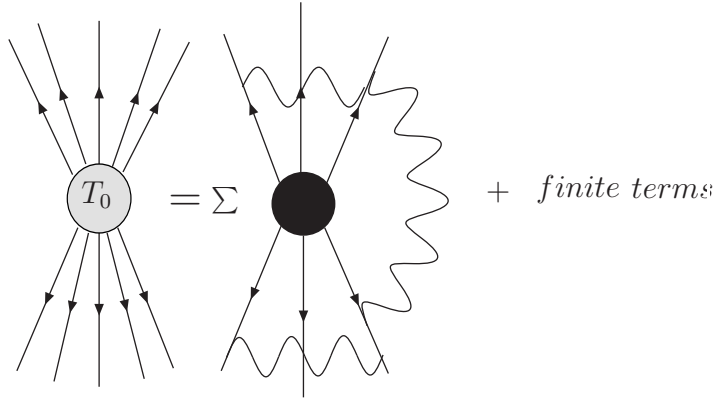
Fig. 5: Diagramatic expression of the fact that the soft
divergences emerge only through the internal photon lines.

\

For each virtual photon attached to two external electron lines
(Fig. 5) we obtain the following expression for the amplitude \ba
\frac{e}{2}\int\frac{d^{4}k}{(2\pi)^{4}i}\overline{u}(p_{i})\gamma_{\mu}\frac{1}{m-p_{i}\!\!\!\!/+k\!\!\!/}M_{0}\frac{1}{m-p_{j}\!\!\!\!\!/-q\!\!/}e^{\frac{i}{2}\widetilde{Q}_{ij}\cdot
k}\gamma_{\nu}u(p_{j})\frac{g^{\mu\nu}}{k^{2}} \nonumber\\\approx
\frac{e^{2}}{k^{2}}\int_{V}\frac{d^{k}}{(2\pi)}\frac{1}{k^{2}}\frac{p_{i}\cdot
p_{j}}{(p_{i}\cdot k)(p_{j}\cdot
k)}M_{0}e^{\frac{i}{2}\widetilde{Q}_{ij}\cdot k} \ea where $V$
denotes the soft photon region $\mu\leq k_{0}\leq\Lambda$ and
$t_{0}\equiv\overline{u}(p_{i})M_{0} u(p_{j})$ is the amplitude
corresponding to the core diagram (The dark blob in Fig. 5,
indicates that part of the full diagram having no infrared
divergences and is called the core diagram). The factor
$\frac{1}{2}$ is required since our procedure counts each Feynman
diagram
 twice. The explicit expression for $t_{1}$ and thus $t_{n}$ reads in general as,
\be
t_{1}=t_{0}\int\frac{d^{4}k}{(2\pi)^{4}i}\frac{e^{2}}{k^{2}+i\varepsilon}\frac{1}{2}\sum_{i,j}^{N}\frac{-\eta_{i}\eta_{j}p_{i}\cdot p_{j}}{(p_{i}\cdot k)(p_{j}\cdot k)}e^{\frac{i}{2}\widetilde{\textbf{Q}}_{ij}\cdot\ k}
\ee
and
\be
t_{n}=t_{0}\frac{1}{n!}\left [\int\frac{d^{4}k}{(2\pi)^{4}i}\frac{e^{2}}{k^{2}+i\varepsilon}\frac{1}{2}\sum_{i,j}^{N}\frac{-\eta_{i}\eta_{j}p_{i}\cdot p_{j}}{(p_{i}\cdot k)(p_{j}\cdot k)}e^{\frac{i}{2}\widetilde{\textbf{Q}}_{ij}\cdot\ k}\right] ^{n}
\ee
where the factor $\frac{1}{n!}$ comes as a result of symmetry considerations in attaching the $n$ internal photon to
 external electron lines. From (42), equation (45) leads to
\be T_{0}=t_{0}e^{\frac{g}{2}} \ee where the factor $g$ is defined
as, \be
g=\int\frac{d^{4}k}{(2\pi)^{4}i}\frac{2e^{2}}{k^{2}+i\varepsilon}\sum_{i,j}^{N}\frac{-\eta_{i}\eta_{j}p_{i}\cdot
p_{j}}{(p_{i}\cdot k)(p_{j}\cdot
k)}e^{\frac{i}{2}\widetilde{\textbf{Q}}_{ij}\cdot k} \ee Squaring
$T_{0}$, we have, \be |T_{0}|^{2}=|t_{0}|^{\ 2}e^{Re\,g} \ee Using
the fact that, \be
\frac{1}{k^{2}+i\varepsilon}=\frac{P}{k^{2}}-i\pi \delta(k^{2})
\ee with $P$ indicating the principal value, we have, \ba &Re \,g&
=\int\frac{d^{4}k}{(2\pi)}2e^{2}\pi\delta(k^{2})\sum_{i,j}^{N}\frac{\eta_{i}\eta_{j}p_{i}\cdot
p_{j}}{(p_{i}\cdot k)(p_{j}\cdot
k)}\sin(\frac{1}{2}\widetilde{\textbf{Q}}_{ij}\cdot \textbf{k})
\nonumber\\
&&=-\int_{\mu}^{\Lambda}\frac{dk_{0}}{k_{0}}\sin(\frac{1}{2}\widetilde{\textbf{Q}}_{ij}\cdot
k_{0})f\approx -f\ln(\frac{\Lambda}{\mu}) \ea where $f$ is as
defined by Eq.(39), so \be
|T_{0}|^{2}=|t_{0}|^{2}\left(\frac{\mu}{\mu}\right)^{f} \ee Now
with inserting $|T_{0}|^{2}$ into the equation (41), we confirm
the finiteness of the physical transition rate as $\mu\rightarrow
0$, \be \omega=|t_{0}|^{2}\left(\frac{\mu}{\Lambda}
\right)^{f}\left(\frac{E_{l}}{\mu}
\right)^{f}=|t_{0}|^{2}\left(\frac{E_{l}}{\Lambda} \right)^{f} \ee
This cancellation  can also be checked order by order. Note that
in the above argument we have confined ourselves
 to the soft photon
region in which the non-commutative IR divergence disappears.
Note also that the diagrams involving triple photon emission have
not been dealt with here. Such processes are completely novel and
lead to
 collinear divergences whose cancellation requires
 such general theorems as KLN.  In the next section we turn to this theorem in non-commutative space.

\section{The Kinoshita-Lee-Nauenberg Theorem in Non-Commutative Space}

We showed that in the non-commutative theories with massless
gauge fields as commutative ones, there are both IR and mass
divergences. In some Feynman diagrams with triple photon
splitting, there exists another kind of collinear divergences.
Some methods have been introduced to cancel these divergences in
Yang-Mills theory, for example in QCD, a constructive approach
leading to a soft-finite asymptotic dynamics which produces a set
of asymptotic states with a $S$-matrix operator with soft finite
elements [29]. These states are essential generalizations of the
coherent states. This is a perturbative procedure and has been
developed in analogy with that used in the Abelian case.
Moreover, there is a fundamental quantum mechanical theorem
greatly advanced by the works of Kinoshita \cite{17} and Lee and
Nauenberg \cite{18} (KLN theorem [25]) that provide the standard
theoretical response to the problem of collinear and soft
divergences. For clarity, we assume the problem contains a certain
parameter $\mu$ and the degeneracy occurs in the total
Hamiltonian only when $\mu \rightarrow 0 $ [25]. For example in
non-commutative QED  $\mu$ can be  mass of the photon.

 {\sl As
they have shown, the occurrence of such divergences are
consequence of high degree of degeneracy in the system. The
cancellation of these divergences can be established without any
explicit use of Feynman graghs, nor even the explicit form of the
Hamiltonian [17,18,25].}

This theorem and also the coherent state approach are based on
unitarity. In this paper, we will try to prove the
non-commutative version of KLN theorem on the condition that
unitarity is conserved i.e. $\theta_{i0}=0$. This theorem reads
as follows :

\textit{In a non-commutative field theory with only space
non-commutativity with massless fields, transition rates are free
of the soft and collinear divergence if the summation over the
initial and final degenerate states is carried out.}

 The
degenerate states means, for example, the states with an electron
and soft photon belonging to the same energy eigenstate as that
of a single electron in the limit of vanishing photon energy. The
proof of non-commutative version of KLN theorem is
straightforward. We begin by investigating the structure of
$S$-matrix in the interaction picture on noncommutative space. In
the interaction picture, the Schr\"{o}dinger equation for the
state $|\alpha,t\rangle$ in the natural unit system is, \be
i\frac{\partial}{\partial t}|\alpha,t \rangle =g\widehat{H}^{\
(\theta)}_{I}|\alpha,t \rangle \ee \be \widehat{H}^{\
(\theta)}_{I}=e^{iH_{0}t}H^{( \theta)}_{I}e^{-iH_{0}t} \ee where
$H_{0}$ is the free Hamiltonian and $H^{\ \theta}_{I}$ is the
interaction term in which the common product between fields has
been replaced by $\star$-product. For example, in the field
theory context $gH^{\,(\theta)}_{I}$ can be of the forms $e\int
d^{3}x \gamma^{\mu}\overline{\psi}\star A_{\mu}\star \psi$.
Defining a unitary operator \be U(t,t_{0})=T\{
\exp(-i\int_{t_{0}}^{t}d\lambda \widehat{H}^{\
(\theta)}_{I}(\lambda))\} \ee where $T$ denotes the time-order
product, the time evolution of the state $|\alpha,t\rangle $ can
be described as \be |\alpha,t\rangle=
U(t,t_{0})|\alpha,t_{0}\rangle \ee Also the $U(t,t_{0})$
satisfies the same differential equation, \be
i\frac{\partial}{\partial t}U(t,t_{0}) =g \widehat{H}^{\
(\theta)}_{I}U(t,t_{0}) \ee Now with the initial condition
$U(t_{0},t_{0})=1$,  equation (57) can be solved to obtain the
$S$-matrix which is defined as [30], \ba
 &S&=U(-\infty,\infty)
  \nonumber\\ &&=\sum_{n=0}^{\infty}\frac{(-i)^{n}}{n!}\int_{-\infty}^{\infty}dt_{1}\int_{-\infty}^{\infty}dt_{2}...\int_{-\infty}^{\infty}dt_{n}
     T \left\{\widehat{H}^{\ (\theta)}_{I}(t_{1})\widehat{H}^{\ (\theta)}_{I}(t_{2})...\widehat{H}^{\ (\theta)}_{I}(t_{n})\right\}
  \nonumber\\ &&=T\left\{\exp(-i\int dt \widehat{H}^{\ (\theta)}_{I}(t))\right\}
\ea

Providing that $\theta_{i0}=0$, the $S$-matrix is unitary:
$S^{\dag}S=SS^{\dag}=1$. So also, we have \be
S=U(-\infty,0)U(0,\infty)=U^{\dag}(0,-\infty)U(0,\infty) \ee So
for the transition rate $|\langle f|S|i\rangle|^{2}$, we get \be
|\langle f|U^{\dag}(0,-\infty)U(0,\infty)|i\rangle|^{2} \ee In
order to see the origin of the IR divergence from the initial and
final degenerate states explicitly,
 we separate the transition rate
into two parts: one including  only the final state $|f\rangle$ and the other including the initial state $|i\rangle$,
\be
|\langle f|U^{\dag}(0,-\infty)U(0,+\infty)|i\rangle|^{2}=\sum_{m}\sum_{n}(R^{\ +}_{mn})^{\ast}R^{\ -}_{mn}
\ee
where
\be
R^{\ +}_{mn}=\langle m|U(0,+\infty)|f\rangle^{\ast}\langle n|U(0,+\infty)|f\rangle
\ee
\be
R^{\ -}_{mn}=\langle m|U(0,-\infty)|i\rangle^{\ast}\langle n|U(0,-\infty)|i\rangle
\ee

For problems in field theory, one may imagine (61) as cutting
rules. Now let us examine the problem by expanding $U$ up to order
$g$ and performing the time integration, from which we will have
 \be
 R^{\ +}_{mn}=\delta_{mf}\delta_{nf}-\frac{g\langle m|H_{I}^{(\theta)}|f\rangle^{\ast}}
 {E_{m}-E_{f}-i\varepsilon}\delta_{nf}-\frac{g\langle n|H_{I}^{(\theta)}|f\rangle}{E_{n}-E_{f}+i\varepsilon}\delta_{mf}
 \ee
 \be
 R^{\ -}_{mn}=\delta_{mi}\delta_{ni}-\frac{g\langle m|H_{I}^{(\theta)}|i\rangle^{\ast}}
 {E_{m}-E_{i}+i\varepsilon}\delta_{nf}-\frac{g\langle n|H_{I}^{(\theta)}|f\rangle}{E_{n}-E_{i}-i\varepsilon}\delta_{mi}
 \ee
 where $E_{r}$ is the eigenvalue of $H_{0}$ for the state acting on state $|r\rangle$. In an example of scattering, the states
 $|m\rangle$ and $|f\rangle$ in $R^{+}$ may be an electron state $|e\rangle$ and the state including an electron
  and an emitting
 photon $|e\,\gamma\rangle$, respectively. If the emitting photon is soft or collinear,
 the states $|e\rangle$ and $|e\,\gamma\rangle$ will have the same energy eigenvalues and are degenerate.
 So in (64) and (65) the denominators will be divergent and we face  an infrared divergence. Suppose
 $D(E)$ is a subspace of the Hilbert space with all degenerate states with energy $E$ ($\mu \rightarrow
 0$).

We may consider the case of soft divergence in non-commutative
QED. So the parameter $\mu$ can be the photon mass and $D(E)$ can
 be consisted of any photon whose energy is $< E_{l}$. Then
 in the quantity

 \be
 \sum_{m\epsilon D(E)}\sum_{n\epsilon D(E)}|\langle m|S|n\rangle|^{\ 2},
 \ee
 the IR divergences are absent. To prove this from (61), it is sufficient to show that $\sum_{r\epsilon D(E)}R^{\pm}_{mn}$ is free from the IR
 divergence. The proof is easy and to order $g$ and all orders of $\theta$ the non-commutative parameter is given from
\ba
    &\sum R^{\pm}_{mn}&= 0 \,\ \ if \ \ m,n \ \ \epsilon \!\!\!/ \ \ D(E)\nonumber\\ &&
                                   = -\frac{g\langle m|H^{(\theta)}_{I}|n\rangle^{\ast}}{E_{m}-E\mp i\varepsilon} \ \ if \ \ m  \ \ \epsilon\!\!\!/ \ \ D(E) \ \ and \ \ n \ \ \epsilon \ \ D(E)\nonumber\\ &&
                                     =-\frac{g\langle n|H^{(\theta)}_{I}|m\rangle  }{E_{n}-E\pm i\varepsilon}\ \  if \ \ m \ \ \epsilon \ \ D(E)\ \ and \ \ n \ \ \epsilon\!\!\!/ \ \ D(E)\nonumber\\ &&
                                     =0 \ \ if \ \ m,n \ \ \epsilon  \ \ D(E)
\ea

Hence $R^{\pm}_{mn}$ (and consequently Eq. (66)) are finite up to
order g. The generalization of this proof to all orders of $g$ is
not difficult. It should be noted that in the above we have
essentially used the unitarity of the operator $U$; i.e., the
condition $\theta_{i0}=0 $. The non-commutative KLN theorem
 guarantees the cancellation of both collinear and soft divergences at each order of
perturbation theory.

\section{Conclusion}
In this paper, the possibility of the non-commutative IR divergence cancellation at one-loop order in physical
cross sections was studied. We defined a non-commutative soft photon which the detector could not see in the
 non-commutative scales. It has been
demonstrated that in a scattering process computed up to one-loop order, if we consider the non-commutative
 soft photon emission,
the non-commutative logarithmic IR divergence in the vertex
correction will be cancelled in the cross section. However, there
are additional non-Abelian type diagrams in which their
non-commutative logarithmic and quadratic divergences can not be
canceled out using the cross section method. This
non-cancellation is attributed to an important difference between
soft and non-commutative IR divergences. The soft divergences are
associated with the classical limit but non-commutative IR
divergences are completely a quantum mechanical effect. We have
further shown the soft divergences that only appear in the
non-planar vertex correction to be cancelled out in the physical
cross section to all orders. To prove this cancellation, the
Weinberg's method was used. In an all order treatment, we also
have examined the diagrams involving the triple photon splitting
that led to collinear divergences similar to Yang-Mills theories.
Some general theorems such as KLN  guarantee that the transition
rates are free from this kind of divergence if we sum over
initial and final degenerate states. In this paper, we studied
the KLN theorem in the non-commutative space ($\theta_{i0}$=0).
According to this theorem, the non-commutative QED will be free
of collinear divergence.



\end{document}